\documentclass[twoside,11pt,preprint]{article}

\usepackage{xspace}
\usepackage{jmlr2e}
\usepackage{subfigure}
\newcommand{\papertitle}{\texttt{secml-malware}: Pentesting Windows Malware Classifiers with Adversarial EXEmples in Python}
\newcommand{\shortpapertitle}{\texttt{secml-malware}: Pentesting EXE Malware Classifiers}
\newcommand{\secmlmalware}{\texttt{secml-malware}\xspace}
\newcommand{\secml}{\texttt{secml}\xspace}
\newcommand{\GAMMA}{GAMMA\xspace}

\newcommand{\myparagraph}[1]{\vspace*{0.05cm} \noindent \textbf{#1}}
\newcommand{\ie}{i.e.\xspace}
\newcommand{\eg}{e.g.\xspace}

\newcommand{\soa}{state-of-the-art\xspace}
\usepackage[table,xcdraw]{xcolor}

\jmlrheading{1}{2024}{1-48}{4/00}{10/00}{meila00a}{Luca Demetrio, Battista Biggio}

\ShortHeadings{\shortpapertitle}{Demetrio and Biggio}
\firstpageno{1}

\begin{document}

\title{\papertitle}

\author{\name Luca Demetrio \email luca.demetrio@unige.it \\
       Università degli Studi di Genova\\
       \AND
       \name Battista Biggio \email battista.biggio@unica.it \\
       Università degli Studi di Cagliari}


\maketitle

\begin{abstract}
Machine learning has been increasingly used as a first line of defense for  Windows malware detection. 
However, recent work has shown that these detectors can be evaded by adversarial EXEmples, carefully-perturbed input malware samples, thus demanding for tools that can ease and automate the adversarial robustness evaluation of such detectors.
Thus, we present \secmlmalware, the first Python library for computing adversarial attacks on Windows malware detectors.
\secmlmalware implements state-of-the-art white-box and black-box attacks on Windows malware classifiers, leveraging a set of manipulations that can be applied to Windows programs while preserving their functionality.
The library can be used to perform the penetration testing and assessment of the adversarial robustness of Windows malware detectors, and it can be easily extended to include novel attack strategies. 
Our library is available at \url{https://github.com/pralab/secml_malware}.

\end{abstract}


\section{Introduction}

Machine learning is extensively used as a first line of defence against the spread of Windows malware.
Both industry and academia are developing increasingly-sophisticated algorithms for extracting malicious patterns from data, leveraging different feature sets and model architectures~\citep{raff2018malware, coull2019activation, saxe2015deep, anderson2018ember}.
Meanwhile, recent work has shown that these learning-based malware detectors can be misled, enabling the attacker to  infect the target device with malware~\citep{demetrio2019explaining, demetrio2020adversarial, demetrio2020efficient, kreuk2018deceiving, suciu2019exploring, sharif2019optimization, anderson2017evading, castro2019aimed}.
The latter can be achieved by applying \emph{practical manipulations} that do not alter the functionality of a malicious program, but rather its file structure.
In this way, the attacker crafts an adversarial EXEmple, \ie, an adversarial example for Windows malware detectors~\citep{demetrio2020adversarial}, which can be run on the target machine even after being manipulated.
Hence, there is the need of open-source tools to test classifiers and defenses against these threats, in order to understand how to mitigate such them, while being one-step ahead of possible attackers.

For this reason, we propose \secmlmalware, the first Python library for creating adversarial EXEmples in input space, providing developers and analysts a tool for performing security evaluations on their machine-learning Windows malware detectors.
The library implements most of the proposed practical manipulations to perturb Windows programs, and it is written on top of the \secml library~\citep{pintor2022secml}.
The structure is modular enough to permit users to code their own attacks and wrap models to test easily.
Also, it is shipped with both white-box and black-box strategies that can be used as-is against Windows malware detectors.
The repository is available at \url{https://github.com/pralab/secml_malware}.
Also, we integrate \secmlmalware as part of \textit{ToucanStrike},\footnote{\url{https://github.com/pralab/toucanstrike}} a command-line tool for creating adversarial EXEmples by typing commands inside the shell. 

\section{\secmlmalware: Architecture and Implementation}

\begin{table}

\begin{minipage}{0.49\textwidth}
\resizebox{\textwidth}{!}{%
\begin{tabular}{cc}
	\rowcolor[HTML]{EFEFEF} 
	\multicolumn{2}{c}{\cellcolor[HTML]{EFEFEF}\textbf{White-box attacks }}                             \\ \hline
	\rowcolor[HTML]{C0C0C0} 
	\multicolumn{1}{c|}{\cellcolor[HTML]{C0C0C0}\textbf{Proposed by}} & \textbf{Practical manipulation} \\ \hline
	\multicolumn{1}{c|}{\cite{demetrio2019explaining}}                & partial dos                     \\ \hline
	\multicolumn{1}{c|}{\cite{demetrio2020adversarial}}               & full dos                        \\ \hline
	\multicolumn{1}{c|}{\cite{demetrio2020adversarial}}               & extend                          \\ \hline
	\multicolumn{1}{c|}{\cite{demetrio2020adversarial}}               & shift                           \\ \hline
	\multicolumn{1}{c|}{\cite{kolosnjaji2018adversarial}}             & padding                         \\ \hline
	\multicolumn{1}{c|}{\cite{kreuk2018deceiving}}                    & slack+padding                   \\ \hline
	\multicolumn{1}{c|}{\cite{suciu2019exploring}}                    & slack+padding                   \\ \hline
	\end{tabular}}

\end{minipage}
\quad
\begin{minipage}{0.49\textwidth}
\resizebox{\textwidth}{!}{%
\begin{tabular}{cc}
	\rowcolor[HTML]{EFEFEF} 
	\multicolumn{2}{c}{\cellcolor[HTML]{EFEFEF}\textbf{Black-box attacks }}                             \\ \hline
	\rowcolor[HTML]{C0C0C0} 
	\multicolumn{1}{c|}{\cellcolor[HTML]{C0C0C0}\textbf{Proposed by}} & \textbf{Practical manipulation} \\ \hline
	\multicolumn{1}{c|}{\cite{demetrio2020efficient}}                 & GAMMA padding                   \\ \hline
	\multicolumn{1}{c|}{\cite{demetrio2020efficient}}                 & GAMMA section inj.              \\ \hline
	\multicolumn{1}{c|}{\cite{demetrio2020adversarial}}               & partial dos                     \\ \hline
	\multicolumn{1}{c|}{\cite{demetrio2020adversarial}}               & full dos                        \\ \hline
	\multicolumn{1}{c|}{\cite{demetrio2020adversarial}}               & extend                          \\ \hline
	\multicolumn{1}{c|}{\cite{demetrio2020adversarial}}               & shift                           \\ \hline
	\multicolumn{1}{c|}{\cite{demetrio2020adversarial}}               & padding                         \\ \hline
	\multicolumn{1}{c|}{\cite{demetrio2020adversarial}}               & slack+padding                   \\ \hline
	\end{tabular}}

\end{minipage}
\caption{Attacks implemented in \secmlmalware, along with the manipulations they apply.}
\label{table:attacks}
\end{table}

The library is divided in three main modules: \texttt{attack}, \texttt{models}, and \texttt{utils}.
Each of these packages is provided with unit tests that asserts the correct behaviour of these techniques.

\myparagraph{The \texttt{attack} module.}
\label{subsec:attack}
This module contains all the attacking strategies, divided in \texttt{whitebox} and \texttt{blackbox} modules, and we provide a complete list of them in Table~\ref{table:attacks}.
Attacks inside the \texttt{whitebox} sub-module are implemented by leveraging practical manipulations that address the perturbing of single bytes inside the program~\citep{demetrio2019explaining, demetrio2020adversarial, kreuk2018deceiving, suciu2019exploring}.
These attacks exploit the ambiguity of the file format, by filling unused space inside the binary or injecting new content in particular positions, but always preserving the original functionality of the sampl.
Since we use \secml, the optimizer uses \texttt{pytorch} \citep{paszke2019pytorch}, and all the gradient computations leverage this framework. 
Attacks inside the \texttt{blackbox} sub-module are implemented using DEAP~\citep{DEAP_JMLR2012}, a library for encoding genetic optimizers, and they span from byte-based to more structural manipulations, \eg section and API injection (\cite{demetrio2020efficient, demetrio2020adversarial}).
We also include \GAMMA~\citep{demetrio2020efficient}, that is a black-box attack leveraging the injection of benign content to fool the target detector, by also keeping low the size of the adversarial malware and the number of queries sent.
Since we use \secml, the optimizer leverage the \texttt{pytorch} framework~\citep{paszke2019pytorch} for computing gradients.

\myparagraph{The \texttt{models} module.}
\label{subsec:models}
This sub-module hosts the definition of two \soa classifiers: a deep neural network, called \textit{MalConv}~\citep{raff2018malware}, and a Gradient Boost Decision Tree (GBDT)~\citep{anderson2018ember}.
Both of them are encapsulated in classes that can be passed to the underlying \secml framework for computing the attacks.
The latter is modular enough for including most models from different frameworks, and the end user can leverage this feature to include their custom target.

\myparagraph{The \texttt{utils} module.}
\label{subsec:utils}
This sub-module contains support code to implement practical manipulations, such as functions for keeping the constraints intact.

Lastly, we also provide a detailed description on how to create a custom \texttt{conda} environment,\footnote{\url{https://anaconda.org}} and a Docker container.\footnote{\url{https://www.docker.com}}

\section{Application Example: Evaluating Adversarial Robustness of MalConv}

To show the potential of our library, we apply both white-box and black-box state-of-the-art attacks already coded in \secmlmalware against a deep neural network called MalConv~\citep{raff2018malware}.
The latter is just an example we chose for simplicity, but \secmlmalware provides wrappings also for other classifiers as well (\eg gradient boosting decision trees and general Pytorch neural networks~\citep{demetrio2020adversarial,demetrio2020efficient}
For this example, we choose to test the following strategies.

\myparagraph{Partial DOS.}
This attack leverage the editing of a fraction of the unused DOS header, kept inside compiled binaries for retro-compatibility~\citep{demetrio2019explaining}.

\myparagraph{Extend.}
This attack leverage the extension of the unused DOS header, by shifting the real one by a custom amount, and injecting adversarial content there~\citep{demetrio2020adversarial}.
This is achieved by exploiting the presence of an offset inside the DOS header that instructs the loader where to look for the content of the program.
Then, injected content must be compliant with the constraints imposed by the file format.

\myparagraph{Shift.}
This attack inject new content before the first section, by shifting all the content by the custom amount~\citep{demetrio2020adversarial}.
This manipulation exploits the offsets inside the header of the program that instructs the loader where the sections start inside the binary.
Since all sections must be aligned to a multiple of the file alignment, the content must match this constraint in order to not break the structure.

\myparagraph{Padding.}
This attack leverages the appending of new bytes at the end of the executable~\citep{kolosnjaji2018adversarial}.
Such content is not controlled by the loader, since there are no pointers to such addition inside neither the code and the header.

\myparagraph{GAMMA-padding.}
This attack leverage the \emph{padding} manipulation to inject content extracted from benign software~\citep{demetrio2020efficient}, whose size is controlled by a regularization parameter to avoid uncontrolled growht of the manipulation size.

\subsection{Experimental results}
We test all the presented strategies in both white-box and black-box settings, aside for \emph{GAMMA-padding} which is evaluated only on the latter.
The results are shown in  Table~\ref{table:malconv_results}.

\noindent \myparagraph{White-box attacks.} We set the maximum number of iterations to 50, and we observe how the detection rate decreases while optimizing the injected content.
Our library is able to identify a weakness to attacks that perturb the header of programs; in particular, the \emph{Extend} attack is able to decrease the detection rate close to 0 in only few iterations.

\noindent \myparagraph{Black-box attacks.} We bound the maximum number of queries to 500 for the black-box attacks.
For \emph{GAMMA-padding}, we extracted 100 \texttt{.data} sections from legitimate programs, and we set the regularization parameter to $10^{-5}$.
Since black-box attacks do not rely on gradients, but they query the model to estimate a direction, they are less effective than their white-box counterparts.
Differently, \emph{GAMMA-padding} it optimizes directly the injection of larger chunks of benign content into the malware sample, rather than trying to optimize each single injected byte.
This is enough for decreasing the detection rate close to 0.1.

\begin{table}[]
\resizebox{\textwidth}{!}{
\begin{tabular}{ccccccccccc}
\multicolumn{11}{c}{\textbf{MalConv original DR: 100\%}}                                                                                                                                                                                                                                                                                                                                                                                                                          \\
                                 & \multicolumn{4}{c}{\cellcolor[HTML]{EFEFEF}\textbf{White-box attacks}}                                                                                                     &                                     & \multicolumn{5}{c}{\cellcolor[HTML]{EFEFEF}\textbf{Black-box attacks}}                                                                                                                                                      \\
                                 & \cellcolor[HTML]{C0C0C0}\textbf{Partial DOS} & \cellcolor[HTML]{C0C0C0}\textbf{Extend} & \cellcolor[HTML]{C0C0C0}\textbf{Shift} & \cellcolor[HTML]{C0C0C0}\textbf{Padding} &                                     & \cellcolor[HTML]{C0C0C0}\textbf{Partial DOS} & \cellcolor[HTML]{C0C0C0}\textbf{Extend} & \cellcolor[HTML]{C0C0C0}\textbf{Shift} & \cellcolor[HTML]{C0C0C0}\textbf{Padding} & \cellcolor[HTML]{C0C0C0}\textbf{GAMMA-padding} \\ \cline{2-5} \cline{7-11} 
\cellcolor[HTML]{C0C0C0}1 iter.  & 60\%                                         & 5\%                                     & 87.5\%                                 & 85\%                                     & \cellcolor[HTML]{C0C0C0}10 queries  & 69\%                                         & 34\%                                    & 80\%                                   & 100\%                                      & 14\%                                           \\ \hline
\cellcolor[HTML]{C0C0C0}25 iter. & 28\%                                         & 5\%                                     & 80\%                                   & 45\%                                     & \cellcolor[HTML]{C0C0C0}250 queries & 56\%                                         & 25\%                                    & 79\%                                   & 100\%                                      & 13\%                                           \\ \hline
\cellcolor[HTML]{C0C0C0}50 iter. & 28\%                                         & 5\%                                     & 80\%                                   & 45\%                                     & \cellcolor[HTML]{C0C0C0}500 queries & 42\%                                         & 10\%                                    & 65\%                                   & 100\%                                      & 12\%                                           \\ \hline
\end{tabular}}
\caption{Detection Rates (DRs) of MalConv against white-box/black-box attacks, optimized with an increasing number of iterations/queries.}
\label{table:malconv_results}
\end{table}

\section{Conclusions and Future Work}
We present \secmlmalware, a tool for pentesting the robustness of machine learning Windows malware classifiers.
To showcase its effectiveness, we present results against a state-of-the-art deep neural network, in both white-box and black-box settings.
We remark this is the first library that contains both gradient and gradient-free techniques focused on this domain, and we have evidence that our library is being employed in research work published in top-tier venues~\citep{quiring2020against, yuste2022optimization, trizna2022quo, rigaki2023stealing, gibert2023certified, kuppa2021adversarial, liu2024defend}.
Currently, \secmlmalware has 195 stars on \textit{GitHub}, 46 forks, and an average monthly quota of 500 downloads.\footnote{\url{https://pypistats.org/packages/secml-malware}}
We have already closed issues raised by the community, by fixing bugs and replying to curiosities from interested users.\footnote{\url{https://github.com/pralab/secml_malware/issues?q=is\%3Aissue+is\%3Aclosed}}
As future work, we plan to extend the attacks implemented in \secmlmalware to also target classifiers that extract information from runtime behaviour of malware.
This line of work would be indeed beneficial also for the defense side, as \secmlmalware would become an ubiquitous tool for testing any kind of machine-learning malware classifier.

\acks{This work has been supported by the European Union's Horizon Europe research and innovation program under the project ELSA, grant agreement no. 101070617; by project SERICS (PE00000014) and FAIR (PE00000013) under the MUR NRRP funded by the EU–NGEU; and by Fondazione di Sardegna under the project ``TrustML: Towards Machine Learning that Humans Can Trust’’, CUP: F73C22001320007.}


\vskip 0.2in

\bibliography{bibliography}

\end{document}